\documentclass[aps,twocolumn,showpacs,amssymb]{revtex4-1}
\date{\today}
\usepackage{amsmath}
\usepackage{amsfonts}
\usepackage{amssymb}

\begin{document}

\title{Decoupling of valence nucleons in configuration-interaction shell model approach}

\author{Chong Qi}
\email{chongq@kth.se}
\affiliation{KTH (Royal Institute of Technology), Alba Nova University Center,
SE-10691 Stockholm, Sweden}
\begin{abstract}
The unusual ratio of the neutron and proton transition matrix elements, $M_n/M_p$, may indicate the exotic feature of decoupled neutron and proton density distributions. In this contribution we show that these transition matrix elements are not the same as the ones calculated within the configuration-interaction shell model approach.
\end{abstract}

\maketitle

Neutron-rich nuclei with large $N/Z$ ratios may exhibit the unusual feature of decoupled neutron and proton density distributions. Such decoupled behavior exists in some typical halo nuclei as well as in heavy carbon and boron
isotopes~\cite{Ogawa03,Dom05,Im04,Ele04,Ele09}. The ratio of the neutron and proton transition matrix elements, $M_n/M_p$, has been used to indicate
differences between neutron and proton distributions,
in comparison with the isoscalar value of $M_n/M_p\sim N/Z$~\cite{Bern79}.

The proton transition matrix element $M_p$ is related to the reduced electric quadrupole transition probability $B(E2)$ as
\begin{eqnarray}\label{exp}
M^2_p=B(E2;0^+_{gs}\rightarrow 2^+_1)/e^2.
\end{eqnarray}
On the other hand, the transition matrix elements can be related to the deformation lengths as~\cite{Ele08,Ele09}
\begin{eqnarray}
M^2_p&=&\left(\frac{3}{4\pi}Z\delta_pR\right)^2,
\end{eqnarray}
and 
\begin{eqnarray}
M^2_n&=&\left(\frac{3}{4\pi}Z\delta_nR\right)^2,
\end{eqnarray}
where $R=1.2A^{1/3}$ is the nucleus radius. $\delta$ is the deformation length which can be evaluated from the inelastic scattering cross sectio measurement. 

In the configuration-interaction shell model, the $B(E2)$ value is calculated as,
\begin{eqnarray}\label{sm}
B(E2;J_i\rightarrow J_f)=\frac{1}{2J_i+1}|e^n_{eff}\mathcal{M}_n+e^p_{eff}\mathcal{M}_p|^2,
\end{eqnarray}
where $e_{eff}$ is the effective charge and $\mathcal{M}$ is the effective transition matrix element calculated from the shell model. These two quantities depend on the choice of the model space.
In the following, we restrict $J_i$ and $J_f$ as the ground and first $2^+$ states, respectively.

In some works~\cite{Ele09,Sag04}, shell model calculations are employed in evaluating the transition matrix elements $M_p$ and $M_n$. However, by comparing Eq. ~(\ref{sm}) with Eq.~(\ref{exp}), one can immediately realize that the two proton transition matrix elements $M_p$ and $\mathcal{M_p}$ are different. They are related to each other by,
\begin{eqnarray}
|eM_p|=|e^n_{eff}\mathcal{M}_n+e^p_{eff}\mathcal{M}_p|.
\end{eqnarray}
These two quantities coincide only when $e^p_{eff}=1$ and $e^n_{eff}=0$ which is not the case in usual shell model calculations.

One may wondering how the neutron transition matrix element $M_n$ can be evaluated in the shell model approach? If we assume isospin symmetry is strictly conserved in the nuclear wave function, for the transition matrix elements of mirror nuclei with exchanged proton and neutron numbers we have the following relation,
\begin{eqnarray}
M_n(N,Z)&=&M_p(Z,N),\\
\mathcal{M}_n(N,Z)&=&\mathcal{M}_p(Z,N).
\end{eqnarray}
Inserting above relations into Eq.~(\ref{sm}), one can can the expression for the neutron transition matrix element as,
\begin{eqnarray}
|eM_n|=|e^p_{eff}\mathcal{M}_n+e^n_{eff}\mathcal{M}_p|.
\end{eqnarray}

The ratio between neutron and proton transition matrix elements can be given as,
\begin{eqnarray}
\frac{M_n}{M_p}=\frac{e^p_{eff}\mathcal{M}_n+e^n_{eff}\mathcal{M}_p}{e^n_{eff}\mathcal{M}_n+e^p_{eff}\mathcal{M}_p}.
\end{eqnarray}
This has immediate effect on our conclusion of the decoupling behavior of protons and neutrons with shell model calculations. For example, shell model calculations in the $psd$ shell~\cite{Qi08} give a ratio of
\begin{eqnarray}
\frac{\mathcal{M}_n}{\mathcal{M}_p}\sim4,
\end{eqnarray}
for the effective transition matrix elements. With the standard effective charges of $e^p_{eff}=1.5e$ and $e^n_{eff}=0.5e$, one get
\begin{eqnarray}
\frac{M_n}{M_p}\sim 2,
\end{eqnarray}
which is different with the ratio $
\frac{\mathcal{M}_n}{\mathcal{M}_p}$ by a factor of two.

Detailed shell-model calculations on the spectroscopic properties of neutron rich B and C isotopes will be published in the future.

\section*{ACKNOWLEDGMENTS}
This work has been supported by the Swedish Research Council (VR).

\end{document}